\documentclass[prl,twocolumn,aps,superscriptaddress]{revtex4-1}
\PassOptionsToPackage{usenames,dvipsnames}{color}
\usepackage{amsmath,amsfonts,amssymb,amsthm,graphicx,bbm,enumerate,times,color}
 \usepackage{hyperref}

\newtheorem{theorem}{Theorem}
\newtheorem{lemma}[theorem]{Lemma}
\newtheorem{corollary}[theorem]{Corollary}
\newtheorem{definition}[theorem]{Definition}

\definecolor{martin}{rgb}{0,.4,1}

\newcommand{\mc}[1]{\mathcal{#1}}

\newcommand{\mb}[1]{\mathbb{#1}}

\newcommand{\e}{\mathrm{e}}
\newcommand{\ii}{\mathrm{i}}

\newcommand{\tr}{\mathrm{Tr}} %old
\newcommand{\Tr}{\mathrm{Tr}} %new

\newcommand{\id}{\mb{1}}

\newcommand{\one}{\mathbf{1}}
\newcommand{\1}{\mathrm{id}}

\newcommand{\mcL}{\mathcal{L}}
\newcommand{\tmcL}{\tilde{\mathcal{L}}}
\newcommand{\mcR}{\mathcal{R}}

\newcommand{\Z}{\mb{Z}}
\newcommand{\R}{\mb{R}}

\renewcommand{\P}{\mathbb{P}}

\renewcommand{\1}{\id}

\newcommand{\norm}[1]{\left\Vert #1 \right\Vert}

\renewcommand{\d}{\mathrm{d}}

\newcommand{\SOp}[2][\mcL]{#1\left[#2\right]}

\definecolor{mjk}{rgb}{.8,0.1,.1}
\definecolor{jens}{rgb}{0,0.5,.9}
\definecolor{albert}{rgb}{0.77,0.294,.549}

\definecolor{henrik}{rgb}{.5,0,0.5}

\newcommand{\tLA}{\tilde{\mcL}_{\tilde{A}}}
\newcommand{\tL}{\tilde{\mcL}}

\newcommand{\fu}{Dahlem Center for Complex Quantum Systems, Freie Universit{\"a}t Berlin, 14195 Berlin, Germany}

\usepackage{hyperref}
\begin{document}
\title{Emergence of spontaneous symmetry breaking in dissipative lattice systems}

\author{Henrik Wilming} 
\affiliation{\fu}
\author{Michael J. Kastoryano}
\affiliation{NBIA, Niels Bohr Institute, University of Copenhagen, 2100 Copenhagen, Denmark}
\author{Albert H. Werner}
\affiliation{\fu}
\author{Jens Eisert}
\affiliation{\fu}

\begin{abstract}
A cornerstone of the theory of phase transitions is the observation that many-body systems exhibiting a spontaneous symmetry breaking in the thermodynamic
limit generally show extensive fluctuations of an order parameter in large but finite systems. In this work, we introduce the dynamical analogue of such a theory. Specifically, we consider local dissipative dynamics preparing a steady-state of quantum spins on a lattice exhibiting a discrete or continuous symmetry but with extensive fluctuations in a local order parameter. We show that for all such processes satisfying detailed balance, there exist metastable symmetry-breaking states, i.e., states that become stationary in the thermodynamic limit and give a finite value to the order parameter. We give results both for discrete and continuous symmetries and explicitly show how to construct the symmetry-breaking states. Our results show in a simple way that, in large systems, local dissipative dynamics satisfying detailed balance cannot uniquely and efficiently prepare  states with extensive fluctuations with respect to local operators.
We discuss the implications of our results for quantum simulators and dissipative state preparation.
\end{abstract}
\maketitle

One of the backbones of  modern physics is the theory of phase transitions, whereby a phase transition is accompanied by a change of an order parameter reflecting the spontaneous breakdown of a symmetry \cite{Pathria2011}. Although this paradigm has been enriched by the existence of topological phases of matter, there still remains a lot to
be learned about these more conventional types of phase transitions.

Usually, phase transitions are studied from a \emph{kinematic} point of view: While at high temperatures the Gibbs state is unique \cite{Intensive},
 below a critical temperature several thermal states, corresponding to the different symmetry-broken phases, exist in the thermodynamic limit. In systems of finite volume the thermal state at any finite temperature is always unique and order parameters associated with a symmetry of the Hamiltonian vanish due to the corresponding symmetry of the Gibbs state. Nevertheless, phase transitions can be associated with extensive fluctuations of the order parameter and can therefore already be witnessed in finite systems. More concretely, the value of order parameters in symmetry-breaking thermal states in the thermodynamic limit due to infinitesimal symmetry-breaking fields can be lower bounded by the magnitude of  fluctuations  in large but finite volumes without symmetry-breaking fields \cite{Griffiths1966,Dyson1978,Koma1994}. 

Such kinematic results do not say anything about how the different phases of matter are \emph{prepared} by a physical mechanism. In this work, we provide a dynamic picture: we consider the preparation of states with extensive fluctuations of a local order parameter in large volumes by dissipative open-systems dynamics, generated by local Liouvillians fulfilling detailed balance.
We then show that under such conditions there are always symmetry-breaking sequences of metastable states, which converge to steady states in the thermodynamic limit. Furthermore, we  prove that if the Liouvillian commutes with the charge operator generating the symmetry, there exist dissipative Goldstone-modes on top of symmetry-broken steady-states.

Similar results have been shown in the case of ground-states of local Hamiltonians by Koma and Tasaki \cite{Koma1994}: Extensive fluctuations in order parameters in ground-states of local Hamiltonians lead to symmetry-breaking ground-states in the thermodynamic limit. In this work, we extend their results  to the case of open systems dynamics.

Our results show that such dissipative processes  cannot uniquely prepare a state with density fluctuations. In particular, if the target steady-state is a Gibbs-state with a temperature below a symmetry-breaking phase transition, also symmetry-breaking phases will become steady-states in the thermodynamic limit.

\paragraph{Implications for quantum simulations.}

Apart from the interpretation of our results in terms of the theory of phases in many-body systems and dissipative phase-transitions \cite{Eisert2010,Diehl2011},
the findings may also have immediate implications on the feasibility of
Gibbs states preparation, in particular at low temperatures.
A key aim of \emph{quantum simulations} is to explore unknown zero temperature phase diagrams of local Hamiltonians that are beyond the reach of
classical computers \cite{BlochSimulation,Rydberg}. At best, such a quantum simulation can hope to prepare Gibbs states at low temperatures, effectively through some
dissipative process, to infer the zero temperature behaviour. However, the present results constitute an obstacle against such a procedure -- a fact that has thus far largely been overlooked. 

\paragraph{Set-up.} For simplicity, we consider sequences of systems defined on finite cubic
lattices $\Lambda\subset \Z^d$ of increasing volume $L^d$, where we
associate to every point in $\Lambda$ a finite-dimensional quantum-system with Hilbert space $\mc{H}_x$. Our results can, however, also be transferred to other
regular lattices  and all our findings equally well apply to
\emph{fermionic} open systems \cite{Diehl2011,Eisert2010,Kastoryano2013},
both interacting and non-interacting, as the required notions of locality carry over immediately. The total system is then described by the Hilbert space $\mc{H}_\Lambda=\otimes_{x\in\Lambda}\mc{H}_x$.
In the following we will often be concerned with the total magnetisation in $z$-direction on a region $X\subseteq \Lambda$ as measured  by the observable
\begin{equation}
S^z_X := \sum_{x\in X}S^z_{\{x\}}
\end{equation}
as well as its (global) density $S^z_\Lambda / {|\Lambda|}$. If we consider a lattice system of spin-1/2 particles we therefore have $S^z_{\{x\}}=\sigma_{\{x\}}^z/2$.
More generally we refer to operators that are sums over local operators supported around individual lattice sites as \emph{extensive quantities}.

The dissipative time-evolution in the Heisenberg-picture is generated by a local Liouvillian super-operator $\mcL^\Lambda$ 
\begin{equation}
A(t) = \SOp[\e^{t \mcL^\Lambda}]{A}, \quad \mcL^\Lambda = \sum_{x\in\Lambda}\mcL_x^\Lambda,
\end{equation}
where square brackets indicate the action of a super-operator and each $\mcL_x^\Lambda$ acts on an observable $A$ as \cite{Lindblad1976}
\begin{equation}
\SOp[\mcL^\Lambda_x]A = \mathrm{i} [H_x,A] + \sum_i \left(L^i_xA(L^i_x)^\dagger - \frac{1}{2}\left\{L^i_x(L^i_x)^\dagger,A\right\}\right),
\end{equation}
with $\{L_x^i\}$ being the Lindblad operators.
Throughout this work we will assume that the terms $\mcL_x^\Lambda$ modelling the dissipative process are strictly local, i.e., all operators $H_x$ and ${L}^i_x$ are  supported exclusively on a ball $B_r(x)$ of radius $r$ centered around $x$ (w.r.t.\ the standard metric of the lattice). However, our results also carry over to the setting of approximately local Liouvillians. We will always assume periodic boundary conditions and uniformly bounded dynamics: i.e. $\norm{\SOp[\mcL^\Lambda_x]A}\leq b \norm{A}$ for some constant $b>0$ independent of $x$ and $\Lambda$.

A steady-state of the dynamics is any state of the system $\omega$ whose expectation values are time-independent, i.e. satisfy
\begin{equation}
\omega\left(\SOp[\mcL^\Lambda]A\right) = 0,
\end{equation}
for any observable $A$ supported in $\Lambda$. Here, we use the notation $\omega(A)=\tr(\rho_\omega A)$ if $\omega$ is represented by the density matrix $\rho_\omega$. Steady-states play a similar role in open systems as ground-states do in closed systems. If the steady-state is unique, any initial state will eventually converge to it in the infinite-time limit and the expectation value of any observable $A$ will approach $\omega(A)\one$.

The locality of the dynamics ensures that the time evolution of (quasi-)local observables is well defined in the thermodynamic limit via
\begin{equation}
A(t) := \lim_{\Lambda \nearrow \Z^d} \SOp[\e^{\mcL^\Lambda t}] A \,,
\end{equation}
for any state $\omega$ on the algebra of (quasi-)local observables. This can be seen using Lieb-Robinson bounds, which can also be proven for local Liouvillian dynamics \cite{Lieb1972,Poulin2010,Barthel2011,Nachtergaele2011,Kliesch}.

Since we are ultimately interested in the thermodynamic limit, we will restrict our attention to local observables, such as order parameters.
We will mostly be interested in sequences of states, for which the expectation value of any fixed local observable becomes constant over time as we go to the thermodynamic limit.
In other words, the time it takes to reach stationarity from such states diverges with the system size.
\begin{definition}[Metastable states]
We call a sequence of states $\omega_\Lambda$ (one for each volume $\Lambda$) \emph{metastable} if it satisfies
\begin{equation}
	\lim_{\Lambda \nearrow \Z^d}|\omega_\Lambda(\SOp[\mcL^\Lambda]A)| = 0
\end{equation}
for all local operators $A$.
\end{definition}
Importantly, note that we only require  \emph{local} expectation values to be time-independent. However, in the thermodynamic limit, these are also the only ones which we can measure and meaningfully talk about.

\paragraph{Detailed balance.} Apart from the above mentioned assumptions we will almost exclusively consider states $\omega$ for which the Liouvillian is in \emph{detailed balance} (or reversible),
meaning that $\omega(A\SOp[\mcL^\Lambda]B)=\omega(\SOp[\mcL^\Lambda]AB)$ \cite{Kossakowski1977}.
Since trace preservation requires $\SOp\one=0$ for any Liouvillian, the assumption of detailed balance
already implies that $\omega$ is a steady-state. Many of the most important classes of Liouvillians satisfy detailed balance \cite{Davies1974,Kossakowski1977,Kastoryano2014}, such as dynamics describing weak coupling to a thermal bath.
Importantly, this form of detailed balance implies that the dynamics
is \emph{purely dissipative}, i.e., we only consider the dissipative part of the dynamics and neglect any unitary contributions. This is because we are interested in the \emph{preparation} of states and not so much in their free dynamics. For the convenience of the reader, we also show in the appendix how the above notion of detailed balance generalises the classical notion for Markov chains.

Similarly to metastable states, we also define asymptotically reversible states. These are sequences of states which become reversible with respect to the dynamics on local observables in the thermodynamic limit.
In particular, asymptotically reversible states are metastable.
\begin{definition}[Asymptotically reversible states]
Let $\mc{L}^\Lambda$ be a sequence of Liouvillians and $\omega_\Lambda$ a sequence of states. We call $\omega_\Lambda$ \emph{asymptotically reversible} if
\begin{align}
\lim_{\Lambda \nearrow \Z^d} \left|\omega_\Lambda(\SOp[\mc{L}^\Lambda]A B) - \omega_\Lambda(A\SOp[\mc{L}^\Lambda]B)\right| = 0
\end{align}
for any two local operators $A,B$.
\end{definition}%

 \paragraph{Discrete symmetries.}
It is well known that thermal states on large but finite lattice systems exhibit extensive fluctuations in order parameters, e.g., the magnetisation density, below the critical temperature. Associated with such fluctuations
are long-range correlations and the existence of several distinct symmetry-breaking phases in the thermodynamic limit \cite{Dyson1978}.
We will now consider the case of a steady-state with a $\Z_2$-symmetry, such as spin-flip along the $z$-direction.
Our main result shows that finite density fluctuations  of the order parameters at arbitrary large volumes in the steady-state of a reversible Liouvillian imply the existence of at least two additional metastable state, which explicitly break the symmetry.

We will illustrate the proof of these results in a simple example: The classical spin-1/2 Ising model. We will assume that we have a reversible strictly local Liouvillian preparing the Ising model at zero temperature, whose state we write as $\omega = (\omega^{+}+\omega^{-})/2$, where $\omega^{+}$ and $\omega^{-}$ are the states with all spins pointing up or down, respectively.
We will now first write the states $\omega^{\pm}$ in a different way, making use of the fact that $\omega$ has fluctuations in the magnetisation density. Then we show that $\omega^{\pm}$ both have to be metastable. It will be clear from the arguments given that also at non-zero temperature below the phase-transition, there are symmetry-breaking metastable states (of course a lattice-dimension larger than one is needed for this to happen). It is however not clear, whether these symmetry-breaking states correspond exactly to the pure thermodynamic phases described by KMS-states in the thermodynamic limit.

In the following, by some abuse of notation, we identify $A$ with its support and call $|X|$ the cardinality of the set $X$. Therefore $|A|$ denotes the volume of the support of $A$. For convenience we also set $|\Lambda|=N$ in the following and omit $\Lambda$-subscripts on states and operators. Due to the fact that the all-up and all-down states are product-eigenstates of the total magnetisation we have $\omega^{\pm}(S^z A) = \omega^{\pm}(A S^z)=\pm N\omega^{\pm}(A)/2$ and $\omega^\pm(S^z A S^z) = N^2\omega^\pm(A)/4$. Defining
\begin{equation}
\tilde{O}^\pm := \frac{1}{\sqrt{2}}\left(\one \pm \frac{S^z}{\omega((S^z)^2)^{1/2}}\right),
\end{equation}
one finds
\begin{align}
\label{eq:symmetry-breaking_def}
\omega^{\pm}(A) = \omega\left(\tilde{O}^\pm A\tilde{O}^\pm \right).
\end{align}
More generally, the symmetry of $\omega$ under spin-flips together with its fluctuations in $S^z$ alone are sufficient to show that we can use eq.~\eqref{eq:symmetry-breaking_def} as the \emph{definition} of candidate symmetry-breaking states, with non-vanishing magnetisation density: If
\begin{equation}\label{eq:fluctuations}
\omega((S^z)^2) \geq (\frac{1}{2}\mu N)^2
\end{equation}
is satisfied, it follows that
\begin{align}
|\omega^{\pm}(S^z)| &= \left|\frac{1}{2}\left(\frac{\omega(S^z)}{2} + \frac{\omega((S^z)^3)}{\omega((S^z)^2)}\right) \pm \frac{\omega((S^z)^2)}{\omega((S^z)^2)^{1/2}}\right|\nonumber
\\ &\geq \frac{1}{2}\mu N,
\end{align}
since the terms with odd-parity under spin-flips vanish.

For $\omega^{\pm}$ to be metastable, we see from \eqref{eq:symmetry-breaking_def} that  $\omega(S^z\SOp{A}S^z)$
has to grow slower than $N^2$ and that $\omega(S^z\SOp{A})$ has to grow
slower than $N$ as we increase the volume. We will only prove the former as the latter follows by a fully analogous argument.

First we point out that $\omega(S^z A S^z) = \omega(S^z [A,S^z]) + \omega((S^z)^2 A)$. The first term is
clearly of order $N$, since $[A,S^z]$ is at most of order
$|A|$ due to the locality of $S^z$ and $A$. We can therefore neglect this term.
We will now assume that the Liouvillian satisfies a certain \emph{approximate Leibniz-rule} and that it implies the metastability of $\omega^\pm$. In a second step, we will prove this property. Hence, assume for a moment that
\begin{equation}
\label{eq:approx_derivation_1}
\SOp{(S^z)^2A} = \SOp{(S^z)^2}A + (S^z)^2 \SOp A + O(N)\,.
\end{equation}
Combining this with reversibility and stationarity of $\omega$ we obtain
\begin{align}
\omega(\SOp{(S^z)^2} A) &= \omega((S^z)^2 \SOp A) \\ &= -\omega(\SOp{(S^z)^2} A) + O(N)\,.\nonumber
\end{align}
Thus $\omega((S^z)^2\SOp A)=0$ up to order $N$, which finishes the proof. What is left to show is eq.~\eqref{eq:approx_derivation_1}. To do that, define $\tilde{A}$ as the smallest region such that $\SOp A=\SOp[\mcL_{\tilde{A}}] A$, where $\mcL_{\tilde{A}}$ contains only those terms of $\mcL$ that are supported within $\tilde{A}$. We obtain
\begin{align}
\SOp{(S^z)^2A}
&= \SOp[\left(\mcL - \mcL_{\tilde A}\right)]{(S^z)^2}A + \SOp[\mcL_{\tilde A}]{(S^z)^2 A} \\
&= \SOp{(S^z)^2}A + (S^z)^2 \SOp A  \nonumber \\
&\quad + \SOp[\mcL_{\tilde{A}}]{(S^z)^2A} -  \SOp[\mcL_{\tilde{A}}]{(S^z)^2}A - (S^z)^2 \SOp[\mcL_{\tilde{A}}]A,\nonumber
\end{align}
where we have used $\SOp[\left(\mcL - \mcL_{\tilde A}\right)]{XA} = \SOp[\left(\mcL - \mcL_{\tilde A}\right)]{X}A$ for any operator $X$.
Writing $S^z = Q+R$, where $Q$ is supported on the complement of $\tilde{A}$ and $R$ is supported on $\tilde{A}$, we see that the term with $Q^2$ cancels out, as $\SOp[\mcL_{\tilde{A}}]{Q^2X}=Q^2\SOp[\mcL_{\tilde{A}}]{X}$ for arbitrary $X$. The operator norm of the remaining terms are either zero due to $\SOp{\one}=0$ or of order $N$, since $\mcL_{\tilde{A}}$ is of order $|\tilde{A}|$, which only differs from $|A|$ by some constant factor due to the locality of the Liouvillian. This finishes the proof.

Note that the argument works for any local order parameter instead of $S^z$ and does not depend on the local dimension of the lattice-model or on any specific detail of the Liouvillian.
In fact it turns out that the states $\omega^\pm$ are not only metastable, but asymptotically reversible. We will state this result as a general theorem.

\begin{theorem}[Reversibility from fluctuations]
\label{thm:horsch}
Let $\mcL^\Lambda$ be a sequence of local Liouvillians that are reversible with respect to a sequence of states $\omega_\Lambda$, fulfilling eq.~(\ref{eq:fluctuations}) with respect to some extensive quantity. Then the corresponding states $\omega_\Lambda^{\pm}$, defined through eq.~\eqref{eq:symmetry-breaking_def}, are asymptotically reversible and thus metastable.
\end{theorem}
We stress that the theorem holds without any requirement on how the order parameter transforms under some symmetry and applies also to non-translationally invariant order parameters. The transformation properties are only necessary to show that the states $\omega^{\pm}$ are symmetry-breaking. Furthermore the theorem also applies to Liouvillians whose interactions decay as a power-law with exponent $\beta$ provided that $\beta>2d$.
The proof of this general case is completely analogous to the one given above, however some technicalities arise due to the approximate locality and the stronger statement about reversibility. We therefore present it in the appendix.

\paragraph{Time scales.} We can also estimate the scaling of the survival time $t_{\mathrm{eq}}$ of the symmetry-breaking states $\omega_\Lambda^{\pm}$ with the system size. From the fact that the states are symmetry-breaking, we can lower-bound the equilibration time by the time it takes for the order parameter to relax to the steady state value. Using Lieb-Robinson bounds, we find in the case of finite-range interactions (see appendix), that the equilibration time $t_{\mathrm{eq}}$ scales at least as
\begin{equation}
t_{\mathrm{eq}} \geq  c L^{{d}/{d+1}},
\end{equation}
for some constant $c>0$.

\paragraph{Continuous symmetries.}
Let us now turn to  continuous symmetries, where our results can be further strengthened. We now assume the existence of an extensive self-adjoint quantity $C$, which we call charge and generates the symmetry.
Furthermore we assume the existence of extensive order parameters  $O^{(1,2)}_\Lambda$, satisfying the commutation relations
\begin{equation}\label{eq:ord_param}
[C_\Lambda,O^{(1)}_\Lambda]=\ii O^{(2)}_\Lambda,\quad [C_\Lambda,O^{(1)}_\Lambda]=-\ii O^{(2)}_\Lambda.
\end{equation}
The simplest example to keep in mind is again given by ferromagnetism, choosing
$C_{\{x\}}=S^z_{\{x\}}$ and $O^{(1)}_{\{x\}}=S^{(x)}_{\{x\}},O^{(2)}_{\{x\}}=S^{(y)}_{\{x\}}$, but we could also deal, for example, with staggered magnetic fields.
We will from now on consider steady-states $\omega_\Lambda$ represented by density
matrices $\rho_\Lambda$ commuting with the charge, i.e.
\begin{equation}
\label{eq:symmetry}
[\rho_\Lambda, C_\Lambda]=0.
\end{equation}
This implies that the state is not symmetry-breaking:
$\omega_\Lambda(O^{(i)}_\Lambda) = 0$ for $i=1,2$. As previously we now assume that $\omega_\Lambda$ exhibits extensive fluctuations in the order parameters,
\begin{align}
\label{eq:lro}
\omega_\Lambda \left((O^{(1)}_\Lambda)^2\right)=\omega_\Lambda \left((O^{(2)}_\Lambda)^2\right)\geq (\mu o|\Lambda|)^2.
\end{align}

With a construction similar to \eqref{eq:symmetry-breaking_def} in terms of the order parameters $O^{(i)}_\Lambda$,   Koma and Tasaki \cite{Koma1994} constructed a family of states $\{\omega_\Lambda^{(M)}$;\, $M\leq |\Lambda|\}$,
which under the above assumptions are asymptotically symmetry breaking in the sense that
\begin{align}
\omega_\Lambda^{(M)}\left(O^{(2)}_\Lambda\right) &= 0, \\
\lim_{M\rightarrow \infty}\lim_{\Lambda \nearrow \Z^d}\frac{1}{|\Lambda|}\omega_\Lambda^{(M)}\left(O^{(1)}_\Lambda\right) &\geq \sqrt{2}\mu o.
\end{align}
For details of the construction see Theorem~\ref{app:thm:koma-tasaki} in the appendix. As in the case of discrete symmetries, we can hence explicitly
construct a family of symmetry breaking states. Furthermore it is clear that we can ``rotate them around'' using the charge $C_\Lambda$ as a generator of rotations.
We thus obtain a whole $U(1)$-manifold of symmetry-breaking states in the thermodynamic limit.

\begin{theorem}[Metastability of symmetry breaking states]
\label{thm:symmetry-breaking}
Under the assumption of eqs.~\eqref{eq:symmetry} and~\eqref{eq:lro}, let $\mcL^\Lambda$ be a sequence of local Liouvillians that are reversible with respect to $\omega_\Lambda$. 
Then for any $M$, the states $\omega_\Lambda^{(M)}$ are asymptotically reversible and hence metastable.
\end{theorem}
Note, that we require the steady state $\omega_\Lambda$ to be symmetric with respect to the charge instead of the dynamics, which would imply \cite{Albert2014}
\begin{equation}
\label{eq:symm_liouvillian}
\SOp[\mcL^\Lambda]{\left[C_\Lambda,A\right]} = \left[C_\Lambda,\SOp[\mcL^\Lambda] A\right]
\end{equation}
for any observable $A$. If the steady-state of $\mcL^\Lambda$ is unique, however, such a symmetry of the Liouvillian ensures that the steady-state is also symmetric
in the sense of eq.~\eqref{eq:symmetry} and our theorem applies. 

The proof of the theorem \ref{thm:symmetry-breaking} uses the same strategy as the one of Theorem~\ref{thm:horsch} and also generalises to Liouvillians whose interactions decay faster than any polynomial: First we prove an approximate Leibniz-rule similar to eq.~\eqref{eq:approx_derivation_1}, which, together with reversibility, implies the result. The details of the proof are  quite technical and presented in the appendix.

\paragraph{Goldstone-modes.}
In closed systems, Goldstone's theorem shows the existence of spin-waves of arbitrarily small energy above symmetry-broken states if the Hamiltonian locally commutes with the charge \cite{Landau1981}. The physical intuition is that a global rotation of all spins does not cost any energy and a spin-wave with very long wavelengths has a locally almost constant magnetisation. Since the Hamiltonian is local, the energetic cost of such a spin-wave is very low and goes to zero as the wave-length goes to infinity. The analogous intuition holds also true in the case of open systems if the Liouvillian is local and symmetric in the sense of eq.~\eqref{eq:symm_liouvillian}. We give an explicit construction of such dissipative Goldstone-modes in the appendix.

\paragraph{Discussion.}
The properties of local dissipative dynamics, such as  \emph{locality } \cite{Poulin2010}, \emph{mixing times} \cite{Kastoryano2013,LogSobolev,Kastoryano2014} and \emph{stability against perturbations}
\cite{Cubitt2013,Kastoryano2013}, have recently attracted a great deal of interest. These results are mainly motivated by the question of whether such dissipative processes can be used for reliably
storing quantum information in \emph{quantum memories} \cite{Fujii14,Herold,Koenig2013}, performing \emph{computations}  
\cite{DissipationComputing,Timing} and
 \emph{quantum simulations} \cite{Rydberg}, or preparing \emph{topological phases of matter} \cite{Diehl2011,TopologyByDissipation}.
 Here, we have shown that they also give a dynamical view-point on the emergence of spontaneous symmetry breaking: our results show that local dissipative dynamics satisfying detailed balance with respect to a state with extensive fluctuations of an order parameter necessarily also prepares different symmetry-breaking phases in the thermodynamic limit. Thus symmetry-breaking phases are dynamically stabilised by dissipative dynamics in detailed balance.

An important feature of our work is that it shifts the perspective of symmetry breaking phase transitions from properties of Hamiltonians to properties of quantum states. This mindset is similar to recent studies in the field of topological order, where the  emphasis has been put on states described by \emph{tensor networks} and their entanglement structure instead of Hamiltonians \cite{1010.3732,WenPhases,PollmannPhases}.

Our results rely on locality and reversibility (detailed-balance) of the dynamics. While locality is clearly necessary, the role of reversibility is not quite as clear: It is known that with simple non-reversible update rules of an asynchronous cellular automaton, it is possible to have a domain of stability in the phase diagram even though this is impossible for equilibrium statistical mechanics models \cite{bennett1985,grinstein2004}.
It is an open problem whether criticality can be induced robustly with non-reversible dynamics without the simultaneous production of metastable states. 
Finally, it will be interesting to study whether similar results hold for discrete time Markov processes. These would give information about the convergence of Markov chain Monte Carlo algorithms, which are typically in detailed balance and are used in many areas of physics.

\paragraph{Acknowledgements.} This work has been supported by the ERC (TAQ), the EU (RAQUEL, AQuS, SIQS), the COST network, and the Studienstiftung des Deutschen Volkes. MJK was supported by the Carlsberg fond and the Villum foundation.

%\bibliography{ssb_open2.bib}

\appendix

\onecolumngrid
\section{Detailed balance}
\label{app:detailed_balance}
As detailed balance plays an important role for the present work, we briefly explain here how the notion of detailed balance that we use precisely generalises the classical notion of detailed balance.
To do that let $\mcL$ be a Liouvillian in detailed balance with the quantum state $\omega$, i.e. $\omega(\mcL(A)B) = \omega(A\mcL{B})$ for any two bounded operators $A,B$.
For simplicity, let us assume that the dynamics takes place on a finite-dimensional Hilbert-space.  We can decompose $\omega$ into mutually orthogonal pure states $\psi_j$ with associated projection operators $P_j$ and probabilities $p_j$. Then we have
\begin{align}
\omega(\SOp[e^{t\mc L}]{P_i}P_j) = p_j \psi_j(\SOp[\e^{t \mc L}]{P_i}) =: p_j \P(i,j;t), 
\end{align}
where $\P(i,j;t)$ denotes the probability to end up in state $\psi_i$ after time $t$ when having started in state $j$.
From detailed balance we then get (upon integrating)
\begin{align}
p_j \P(i,j;t) = \omega(\SOp[e^{t\mc L}]{P_i}P_j) = \omega(P_i \SOp[e^{t\mc L}]{P_j}) = p_i \P(j,i;t),
\end{align}
which is precisely the condition of detailed balance in a classical Markov chain defined over the states $\psi_j$ with transition probabilities $\P(i,j;t)$.
In particular, if $\omega$ is a Gibbs-state of a non-degenerate Hamiltonian at inverse temperature $\beta$, the states $\psi_j$ are energy-eigenstates associated to energies $E_j$ and we get the well-known relation
\begin{align}
\P(i,j;t) = \e^{-\beta (E_j-E_i)}\P(j,i;t). 
\end{align}

\section{General proof for discrete symmetry breaking}
\label{app:proof1}
In this section we prove Theorem~\ref{thm:horsch} for the general case of approximately local Liouvillians. The essential ideas are the same as in the proof for compactly supported Liouvillians presented in the example of the Ising model, but we have to estimate the corrections due to the fact that the Liouvillians are not compactly supported. Again we always assume periodic boundary conditions for simplicity. Let us first properly define Liouvillians with non-compact support. Then we will precisely formulate the theorem and prove it. Informally, we say that a Liouvillian is appproximately local if each term $\mc{L}^{\Lambda}_x$ may be well approximated by a compactly supported term $\tilde{\mcL}^{\Lambda}_x$ with support in a ball $B_l(x)$ of radius $l$ around $x$. The error is quantified by a function $f$:
\begin{definition}[$f$-local Liouvillian] Let $f:\Z^d\rightarrow \R$ with $f(0)=1$ be given. A sequence of Liouvillians $\mc{L}^\Lambda=\sum_{x\in\Lambda}\mc{L}_x^\Lambda$ is \emph{$f$-local} if there exists a sequence of compactly supported Liouvillians $\tilde{\mc{L}}^\Lambda=\sum_x \tilde{\mcL}^\Lambda_x$ such that
\begin{align}
\norm{\SOp[\mcL^\Lambda_x]{A}-\SOp[\tilde{\mcL}^\Lambda_x]{A}} \leq c\norm{A}f(l),
\end{align}
where $\tilde{\mcL}^\Lambda_x$ is supported within $B_l(x)$ and $b>0$ is a constant.
\end{definition}
%%%%
\begin{definition}[Approximately local Liouvillian] We will say that $\mc{L}^\Lambda$ is \emph{approximately local} if it is $f$-local and $f$ decays at least as fast as
\begin{align}
f(l) = \frac{1}{1+l^\beta},\quad \beta > 2d.
\end{align}
\end{definition}
%%%
Instead of considering the order parameter $S^z$, we will from now on consider an arbitrary order-parameter $O_\Lambda =\sum_{x\in\Lambda} O_{\{x\}}$, where $O_{\{x\}}$ is compactly supported around lattice site $x$ and $\norm{O_{\{x\}}}\leq o$ for all $x\in \Z^d$.  Given a state $\omega_\Lambda$ we define the states
\begin{align}
\omega_\Lambda^\pm(A) := \omega(\tilde{O}^\pm_\Lambda A \tilde{O}^\pm_\Lambda),\quad\text{with}\quad \tilde{O}^\pm := \frac{1}{\sqrt{2}}\left(\one \pm \frac{S^z}{\omega((S^z)^2)^{1/2}}\right).
\end{align}
The precise theorem that we want to prove now is the following.
\begin{theorem}[Reversibility from fluctuations]
Let $\mc{L}^\Lambda$ be an approximately local Liouvillian that is in detailed balance with respect to the sequence of states $\omega_\Lambda$. Assume the existence of a $\Z_2$-symmetry $U_\Lambda$ such that
\begin{align}
\omega_\Lambda(A) = \omega_\Lambda(U_\Lambda A U_\Lambda^{-1}),\quad O_{x} = -U_\Lambda O_{x}U_\Lambda^{-1}
\end{align}
and that there exists a constant $0<\mu<1$ such that
\begin{align}
\omega_\Lambda(O_\Lambda^2) \geq (\mu o |\Lambda|)^2.
\end{align}
Then the states $\omega^\pm_\Lambda$ are asymptotically reversible and hence metastable.
\end{theorem}

Proof: For simplicity, we will drop the $\Lambda$ labels on all operators and states; in particular we will write $O$ instead of $O_\Lambda$ and $\omega$ instead of $\omega_\Lambda$. We will also set $N:=|\Lambda|$.
It will be useful to introduce the following quantity, which measures how far the action of a Liouvillian deviates from a derivation, i.e., fulfils the Leibniz-rule,
\begin{equation}
\Gamma_{\mcL}(X,Y) := \SOp[\mcL]{XY} - \SOp[\mcL]{X}Y - X\SOp[\mcL]{Y}.
\end{equation}
We have to prove that the states $\omega^{\pm}$ are asymptotically reversible, i.e.,
\begin{align}
\Delta(A,B) := \omega^{\pm}(A\SOp[\mcL]{B}) - \omega^{\pm}(\SOp[\mcL]{A}B) \rightarrow 0
\end{align}
as the system size increases. To do that, we will show separately that
\begin{align}
\lim_{\Lambda \nearrow \Z^d} \frac{\omega\left(O^\dagger O (A \SOp[\mcL]{B} - \SOp[\mcL]{A}B)\right)}{|\Lambda|^2} = 0,
\lim_{\Lambda \nearrow \Z^d} \frac{\omega\left(O (A \SOp[\mcL]{B} - \SOp[\mcL]{A}B)\right)}{|\Lambda|} = 0.\nonumber
\end{align}
Let us first show that, due to reversibility, it suffices to show that for any local operators $A,B$ we have
\begin{align}
\label{eq:app:toshow_horsch}
\lim_{\Lambda \nearrow \Z^d}\frac{\omega\left(\Gamma_{\mcL}(O^\dagger O, A)B\right)}{|\Lambda|^2} = 0,\quad \lim_{\Lambda \nearrow \Z^d}\frac{\omega\left(\Gamma_{\mcL}(O, A)B\right)}{|\Lambda|} = 0.
\end{align}
Indeed, suppose the two properties are true. Then we can use reversibility to write
\begin{align}
\frac{\omega\left(O^\dagger O (A\SOp[\mcL]{B}-\SOp[\mcL]{A}B)\right)}{|\Lambda|^2} &=  \frac{\omega\left((\SOp[\mcL]{O^\dagger O A}-\SOp[\mcL]{A})B\right)}{|\Lambda|^2} = \frac{\omega(\Gamma_{\mcL}(O^\dagger O,A)B)}{|\Lambda|^2} + \frac{\omega(\SOp[\mcL]{O^\dagger O}AB)}{|\Lambda|^2}. 
\end{align}
By our assumption \eqref{eq:app:toshow_horsch}, the first term on the right hand side vanishes in the thermodynamic limit and we obtain
\begin{equation}
\lim_{\Lambda \nearrow \Z^d} \frac{\omega\left(O^\dagger O (A\SOp[\mcL]{B}-\SOp[\mcL]{A}B)\right)}{|\Lambda|^2} = \lim_{\Lambda \nearrow \Z^d} \frac{\omega(\SOp[\mcL]{O^\dagger O}AB)}{|\Lambda|^2}
\end{equation}
We will now use two different ways to evaluate this equation.  On the one hand, we can use reversibility to obtain
\begin{equation}
\lim_{\Lambda \nearrow \Z^d} \frac{\omega\left(O^\dagger O (A\SOp[\mcL]{B}-\SOp[\mcL]{A}B)\right)}{|\Lambda|^2} = \lim_{\Lambda \nearrow \Z^d} \frac{\omega(O^\dagger O\SOp[\mcL]{AB})}{|\Lambda|^2}.
\end{equation}
On the other hand, we can write
\begin{align}
\omega(\SOp[\mcL]{O^\dagger O}AB) &= -\omega(\Gamma_{\mcL}(O^\dagger O, AB)) + \omega(\mcL(O^\dagger O AB)) - \omega(O^\dagger O \SOp[\mcL]{AB}) \nonumber\\
&= -\omega(\Gamma_{\mcL}(O^\dagger O, AB)) - \omega(O^\dagger O \SOp[\mcL]{AB}). 
\end{align}
But since $AB$ is also a local operator we obtain from assumption \eqref{eq:app:toshow_horsch} that
\begin{align}
\lim_{\Lambda \nearrow \Z^d} \frac{\omega\left(O^\dagger O (A\SOp[\mcL]{B}-\SOp[\mcL]{A}B)\right)}{|\Lambda|^2} &= - \lim_{\Lambda \nearrow \Z^d}\frac{\omega(\Gamma_{\mcL}(O^\dagger O, AB))}{|\Lambda|^2} -  \lim_{\Lambda \nearrow \Z^d} \frac{\omega(O^\dagger O\SOp[\mcL]{AB})}{|\Lambda|^2}  \\
&= -  \lim_{\Lambda \nearrow \Z^d} \frac{\omega(O^\dagger O\SOp[\mcL]{AB})}{|\Lambda|^2}.\nonumber
\end{align}
In other words we have
\begin{align}
-  \lim_{\Lambda \nearrow \Z^d} \frac{\omega(O^\dagger O\SOp[\mcL]{AB})}{|\Lambda|^2} = \lim_{\Lambda \nearrow \Z^d} \frac{\omega\left(O^\dagger O (A\SOp[\mcL]{B}-\SOp[\mcL]{A}B)\right)}{|\Lambda|^2} =  \lim_{\Lambda \nearrow \Z^d} \frac{\omega(O^\dagger O\SOp[\mcL]{AB})}{|\Lambda|^2}, 
\end{align}
which just means
\begin{align}
\lim_{\Lambda \nearrow \Z^d} \frac{\omega\left(O^\dagger O (A\SOp[\mcL]{B}-\SOp[\mcL]{A}B)\right)}{|\Lambda|^2} = 0.
\end{align}
Essentially the same argument also works to show
\begin{align}
\lim_{\Lambda \nearrow \Z^d} \frac{\omega\left(O (A \SOp[\mcL]{B} - \SOp[\mcL]{A}B)\right)}{|\Lambda|} = 0
\end{align}
and therefore $\lim_{\Lambda \nearrow \Z^d} \Delta (A,B) = 0$.

What is left is to prove the properties given in eq.~\eqref{eq:app:toshow_horsch}. To do that, first we approximate each term $\mcL_x$ in the Liouvillian by a truncated Liouvillian $\tL_x$ that is supported on a ball of radius $L^\alpha$ around $x$, where $0<\alpha<1$ is to be chosen later. By assumption, for each term this introduces an error given by
\begin{eqnarray}
	\norm{\SOp[\mcL_x]{X} - \SOp[\tL_x]{X}} \leq \norm{X}c f(L^\alpha) &\leq&
	\norm{X}c \frac{1}{1+L^{\alpha\beta}}. 
\end{eqnarray}
We will collect the error terms in a Liouvillian $\mcR$, so that $\mcL = \tL + \mcR$.
For any local operator we will denote by $\tLA$ the Liouvillian containing all terms of $\tL$ whose support has overlap with $A$. Denote the support of this Liouvillian by $\tilde{A}$. We can then make use of the following useful Lemma.
\begin{lemma}[Approximate derivation]
\label{lemma:approx_derivation_app}
For any operator $X$, any local operator $A$ and any strictly local Liouvillian $\tL$ we have
\begin{align}
\Gamma_{\tL}(X,A) = \Gamma_{\tLA}(X,A), 
\end{align}
\begin{proof}
This follows immediately from $\SOp[(\tL - \tLA)]{XA}=\SOp[(\tL - \tLA)]{X}A$.
\end{proof}
\end{lemma}
Since $\Gamma$ is linear in the Liouvillian we can write
\begin{equation}
\Gamma_{\mcL}(O^\dagger O,A) = \Gamma_{\mcR}(O^\dagger O,A) + \Gamma_{\tLA}(O^\dagger O,A). 
\end{equation}
By assumption, $\omega(O^\dagger O)$ is of the order $|\Lambda|^2$ and therefore we are done once we can show
\begin{align}
\frac{||\Gamma_{\mcR}(O^\dagger O,A)B||}{L^{2d}}\rightarrow 0, \quad \frac{||\Gamma_{\tLA}(O^\dagger O,A)B||}{L^{2d}}\rightarrow 0,
\end{align}
in the limit $L\rightarrow \infty$. For the first term, using sub-multiplicativity of the norm and the triangle-inequality, we get
\begin{eqnarray}
\frac{\norm{\Gamma_{\mcR}(O^\dagger O,A)B}}{L^{2d}}
&=&
\frac{\norm{\SOp[\tL_{\mcR}]{O^\dagger OA}B - \SOp[\tL_{\mcR}]{O^\dagger O}AB - O^\dagger O \SOp[\tL_{\mcR}]{A}B}}{L^{2d}}\nonumber\\
&\leq&
\frac{  3 |\Lambda|\norm{O^\dagger O} \norm{A} \norm{ B} }{L^{2d}} c f(L^\alpha)\nonumber\\
&\leq&
\frac{  3 o^2 |\Lambda|^3 \norm{A} \norm{ B} }{L^{2d}} c f(L^\alpha),
\end{eqnarray}
making use of $O= \sum_{x\in \Lambda} O_{x}$. Therefore, \begin{equation}
\frac{\norm{\Gamma_{\mcR}(O^\dagger O,A)B}}{L^{2d}}
\leq
  3 o^2   \norm{A} \norm{ B}  c \frac{L^d}{1+L^{\alpha\beta}}.
\end{equation}

%
%\begin{eqnarray}
%&\leq???& 3 o^2 \norm{A}\norm{B}|\Lambda|^3 c \frac{f(l)}{L^{2d}} = c o^2 \norm{A} c \frac{L^d}{1+L^{\alpha\beta}}.
%\end{eqnarray}
Thus, we see that the term vanishes in the thermodynamic limit as long as $\beta > d/\alpha$.
For the second term, we first decompose $O$ as $O=Q+R$, where $Q$ is supported on the complement of $\tilde{A}$ and $R$ is supported on $\tilde{A}$.
Then we have $\Gamma_{\tLA}(Q,X)=Q\Gamma_{\tLA}(\one,X)=0$, since $\Gamma_{\mcL}(\one,X)=0$ for any Liouvillian $\mcL$ and operator $X$. This implies
\begin{equation}
\Gamma_{\tLA}(O^\dagger O,A)B = 2Q\Gamma_{\tLA}(R,A)B + \Gamma_{\tLA}(R^2,A)B.
\end{equation}
Therefore, a norm-estimate gives
\begin{align}
\frac{\norm{\Gamma_{\tLA}(O^\dagger O,A)B}}{L^{2d}} \leq K o^2 \norm{A} \norm{B} \frac{|\tilde{A}|^2}{L^d}   = K o^2 |A| L^{(2\alpha-1)d},
\end{align}
where $K$ is some positive constant. The term thus converges to zero for $\alpha<1/2$. By essentially the same arguments we can bound the quantities
$\Gamma_{\mcR}(O,A)$ and $\Gamma_{\tLA}(O,A)$, which yield the same constraints on $\alpha$ and $\beta$. Concluding, we see the theorem holds true for any $\beta> 2d$.

\subsection{Survival time scale}
In this section we prove the lower bound on the survival time of the symmetry-breaking states. For simplicity,
we will consider finite-range Liouvillians of range $r$ with steady-state $\omega$.
It should be clear, however, that the same argument can also be applied to approximately local Liouvillians with exponentially decaying tails.
The proof will combine our techniques for the proof of metastability with Lieb-Robinson bounds.
From the proof of metastability in the main-text it is clear that
\begin{align}
\label{app:eq:derivative}
\left|\omega^\pm_\Lambda (\SOp[\mcL^\Lambda]{A})\right| \leq k_1 \frac{\norm{A} |\tilde{A}|}{|\Lambda|},
\end{align}
for some constant $k_1>0$ and any local operator $A$. Dissipative Lieb-Robinson bounds \cite{Kliesch} tell us that we can approximate time-evolved local observables by observables which are supported in the finite \emph{Lieb-Robinson cone}. Let $A$ be a local observable, then we denote the time-evolved observable on the volume $\Lambda$ by
\begin{align}
\SOp[\exp(t \mcL^\Lambda)]{A} =: A^\Lambda(t).
\end{align}
Lieb-Robinson bounds are valid for local dissipative systems in a very similar way as
they hold for local Hamiltonian systems \cite{Kliesch}. They give rise to
a Lieb-Robinson velocity $v>0$ that depends only on the dimension
$d$ of the lattice (here chosen to be a cubic lattice) as well as the range $r$ and the strength of the Liouvillian. They can be used to
 show that $A^\Lambda(t)$ can be approximated by an observable $A^{\vee}(t)$
that is supported within a set that only contains lattice-sites at most $\tilde{v} t \leq L$ away from $A$, as long as
$\tilde v> v$, up to an error of approximation that is is exponentially small in $\tilde{v}$.
More specifically,
\begin{align}
\label{app:eq:lr-bounds}
\norm{A^\Lambda (t) - A^\vee(t)} \leq k_2 \norm{A} (\tilde{v} t)^{d-1}\exp(-(\tilde{v}-v)t),
\end{align}
again for a constant $k_2>0$  depending on $d$, $r$ and the norm of the Liouvillian.
Combining this with the previous estimate, we get
\begin{align}
\left|\omega^\pm_\Lambda(\SOp[\mcL]{A^\Lambda(t)-A^\vee(t)})\right| \leq k_1 k_2 \norm{A} (\tilde{v} t)^{d-1}\exp(-(\tilde{v}-v)t).
\end{align}
Notice that the bound is independent of the system size and the right hand side can be made arbitrarily small, uniformly in $t$, by
suitably increasing $\tilde{v}$.
The dependence on the dimension $d$ in this bound is made more explicit in Refs.\ \cite{Kliesch,Kastoryano2013}.

With these ingredients, we now bound the minimal time $t_{\mathrm{eq}}>0$ that it takes to change the expectation value of an \emph{on-site} observable $A$, such as the order-parameter, by a fixed value $\Delta A$. In order to arrive at a bound for this minimal time, we write
\begin{align}
\Delta A < \left|\omega^\pm_\Lambda(A(t_{\mathrm{eq}}) -A(0))\right| &\leq \int_{0}^{t_\mathrm{eq}}\left|\omega^\pm_\Lambda\left(\frac{\d A(s)}{\d s}\right)\right| \d s = \int_{0}^{t_\mathrm{eq}}\left|\omega^\pm_\Lambda\left(\SOp[\mcL^\Lambda]{A^\Lambda (s)}\right)\right| \d s\nonumber \\
&\leq \int_{0}^{t_\mathrm{eq}}\left|\omega^\pm_\Lambda\left(\SOp[\mcL^\Lambda]{A^\vee (s)}\right)\right| \d s + \int_{0}^{t_\mathrm{eq}}\left|\omega^\pm_\Lambda\left(\SOp[\mcL^\Lambda]{A^\Lambda (s)-A^\vee(s)}\right)\right| \d s\nonumber \\
&\leq k_1 \int_{0}^{t_\mathrm{eq}} \frac{\norm{A} ((2l+1)+2\tilde{v}s))^d}{|\Lambda|}\d s +  k_1 k_2  \norm{A} \int_{0}^{t_\mathrm{eq}}   (\tilde{v} s)^{d-1}\exp(-(\tilde{v}-v)s) \d s \nonumber\\
&\leq \norm{A} \left(k_1' \frac{(\tilde{v}t_{\mathrm{eq}})^{d+1}}{|\Lambda|} + k_1 k_2  \delta(\tilde{v},v, d)\right),
\end{align}
for a suitable constant $k_1'>0$ independent of the system size.
Here,
\begin{equation}
	\delta(\tilde{v},v, d):=
	\int_{0}^\infty   (\tilde{v} s)^{d-1}\exp(-(\tilde{v}-v)s) \d s
        = \frac{(d-1)!}{(1-v/\tilde{v})^d}\frac{1}{\tilde{v}}
	>0
\end{equation}	
converges to zero with increasing $\tilde{v}$, and otherwise is dependent on the dimension $d$ and the Lieb-Robinson velocity $v>0$,
but again independent of the system size. For any dimension $d$ and any given local Liouvillian with Lieb-Robinson velocity $v>0$,
one can always choose a $\tilde v>0$ such that
\begin{equation}
	\delta(\tilde{v},v, d)<\frac{\Delta A}{\norm{A}}\frac{1}{k_1 k_2}.
\end{equation}	
Using that $|\Lambda|=L^d$,
it then follows that
\begin{align}
t_{\mathrm{eq}} > \frac{1}{\tilde{v}}\left(\frac{{\Delta A}/{\norm{A}} - k_1 k_2 \delta(\tilde{v},v, d)}{k_1'}\right)^{1/(d+1)} L^{{d}/{d+1}} > c L^{{d}/{d+1}}
\end{align}
for a suitable $c>0$,
which finishes the proof. The restriction to an on-site operator $A$ was made for reasons of simplicity of the argument only, and an analogous analysis holds for any strictly local operator $A$ as well.

\section{Continuous symmetry breaking}
\label{app:proof2}
In this section we consider the case of continuous symmetry breaking and prove a theorem which yields as corollary Theorem \ref{thm:symmetry-breaking} of the main text. Compared to the case of discrete symmetry
breaking, we will have to assume slightly stronger locality properties for the Liouvillian.
\begin{definition}[Short-range Liouvillian] An $f$-local Liouvillian is \emph{short-ranged} if $f$ decays as least as fast as $\exp(- l^\alpha/\xi)$ for some strictly positive constants $\alpha>0$ and $\xi>0$.
\end{definition}

As in the case of discrete symmetry braking, we will consider explicit families of states which are symmetry-breaking in the thermodynamic limit. These families have been introduced by Koma and Tasaki. To simplify their notation let us first introduce a family of functionals on local observables. Let $m,m'$ be integers such that $|m|,|m'|\leq M$. Using the notation from the main-text, we define the functionals
\begin{equation}
\chi^{(m,m')}_\Lambda (A) := \frac{\omega_\Lambda\left( (O_\Lambda^-)^{k'}A(O_\Lambda^+)^k\right)}{Z(k)Z(k')},
\end{equation}
with $Z(k) = \omega_\Lambda\left((O_\Lambda^-)^{k}(O_\Lambda^+)^{k}\right)^{1/2}$. Here we use the shorthand $(O_\Lambda^+)^m=(O_\Lambda^-)^{-m}$ if $m < 0$. 

\begin{theorem}[Symmetry breaking states \cite{Koma1994}] \label{app:thm:koma-tasaki} For any $M<|\Lambda|$ define the states
\begin{align}
\omega_\Lambda^{(M)}(A):= \frac{1}{2M+1}\sum_{k=-M}^M\sum_{k'=-M}^M \chi^{(m,m')}_\Lambda(A),
\end{align}
Assume that the $\omega_\Lambda$ are represented by density matrices commuting with the charge: $[\rho_\Lambda,C_\Lambda]=0$. If the condition
\begin{align}
\omega_\Lambda \left((O^{(1)}_\Lambda)^2\right)=\omega_\Lambda \left((O^{(2)}_\Lambda)^2\right)\geq (\mu o|\Lambda|)^2
\end{align}
is fulfilled, the states $\omega_\Lambda^{(M)}$ are asymptotically symmetry breaking in the sense that
\begin{align}
\omega_\Lambda^{(M)}\left(O^{(2)}_\Lambda\right) &= 0, \\
\lim_{M\rightarrow \infty}\lim_{\Lambda \nearrow \Z^d}\frac{1}{|\Lambda|}\omega_\Lambda^{(M)}\left(O^{(1)}_\Lambda\right) &\geq \sqrt{2}\mu o.
\end{align}
\end{theorem}

In the following we will drop again the $\Lambda$ from all the operators and again set $N=|\Lambda|$ for simplicity of notation. To state our main result about continuous symmetry breaking, we define the quantities
\begin{equation}
\Delta^{(m,m')}(A,B) := \chi^{(m,m')}\left(B\SOp[\mcL]{A}\right)-\chi^{(m,m')}\left(\SOp[\mcL]{B}A\right),
\end{equation}
which measure in how far the functionals $\chi^{(m,m')}$ are reversible with respect to $\mc{L}$. 

\begin{theorem}[Continuous symmetry breaking]
Suppose $\mcL$ is a short-range Liouvillian that satisfies detailed balance with respect to $\omega$. Furthermore suppose that $\omega$ fulfils the
assumptions \eqref{eq:symmetry} and \eqref{eq:lro}. Then
\begin{equation}
\lim_{\Lambda \nearrow \Z^d}|\Delta^{(m,m')}(A,B)| = 0,
\end{equation}
for any two local operators $A,B$.
\end{theorem}

\begin{corollary}[Convergence to a reversible steady state]
Any state obtained from linear combinations of the $\chi^{(m,m')}$ is asymptotically reversible. In particular, the states $\omega^{(M)}$ in the main-text are asymptotically reversible and hence metastable.
\end{corollary}

We will split the proof into several Lemmas. The first Lemma was proven by Koma and Tasaki and will turn out to be essential. The second Lemma makes use of it and let's us rewrite the problem in a way which will enable us to make use of detailed balance.
\begin{lemma}[Koma, Tasaki \cite{Koma1994}]
\label{lemma:koma}
Let \eqref{eq:symmetry} and \eqref{eq:lro} be fulfilled for a state $\omega$ represented by $\rho$. Let $A$ be some finite region and decompose $O^+$ as $O^+ = Q_A + R_A$, where $Q_A$ is supported on the complement of $A$ and $R_A$ is supported on $A$. Then we have the inequalities
\begin{align}
\label{eq:inequality}
\frac{\Tr(Q_A^{m-k}\rho (Q_A^*)^{m-k})}{\Tr(Q_A^{m}\rho (Q_A^*)^{m})} \leq (\mu o N)^{-2k}
\end{align}
and
\begin{equation}
\label{eq:ratio}
r_A^{(M)} = \left|\frac{\tr\left((O^+)^M \rho (O^-)^M\right)}{\Tr(Q_A^{M}\rho (Q_A^*)^{M})}\right| \geq 2 - \exp(\frac{2|A|M}{\mu N}) \geq 2 - \e^{\mu/8}.
\end{equation}
for $N\geq \frac{16|A|^2}{\mu^2}$ and $|\frac{m}{N}|\leq \frac{\mu^2}{16|A|}$. \nonumber
\begin{proof}
We reproduce the proof at the end of the Appendix for the reader's convenience.
\end{proof}
\end{lemma}
\begin{lemma}[Local observables]\label{lemma:local_observable}
Let $A$ be any local observable. Then
\begin{equation}
\left|\tr\left(\chi^{(m,m')} A \right)\right| \leq O\left(\frac{M |A|\norm{A}}{N}\right) + \left|\frac{\tr\left( \rho (O^-)^{m'}(O^+)^{m} A \right)}{\tr\left((O^+)^{m} \rho (O^-)^{m}\right)^{1/2}\tr\left((O^+)^{m'} \rho (O^-)^{m'}\right)^{1/2}}\right|.
\end{equation}
\begin{proof}
First we split the expectation values into
\begin{equation}
\label{eq:splitting}
\tr\left(\rho (O^-)^{m'} A (O^+)^m\right) = \tr\left(\rho (O^-)^{m'} [A,(O^+)^m]\right) + \tr\left(\rho (O^-)^{m'} (O^+)^m A\right).
\end{equation}
We have to show that the first term divided by the denominator is of the corresponding order. To do that let us split up $O^+$ as $O^+ = Q_A + R_A$, where $Q_A$ is supported on the complement of $A$ and $R_A$ is supported on $A$. This implies that $[Q_A,A]=0$ and $[Q_A,R_A]=0$.
Using a binomial expansion we obtain
\begin{align}
\text{1st term} &= \sum_{k=0}^{m'}\sum_{l=0}^m {m' \choose k}{m\choose l}\tr\left(\rho (Q_A^*)^{m'-k}(R_A^*)^k[A,Q_A^{m-l}R_A^{l}]\right) \\
&= \sum_{k=0}^{m'}\sum_{l=1}^m {m' \choose k}{m \choose l}\tr\left(\rho (Q_A^*)^{m'-k}(R_A^*)^k[A,R_A^{l}]Q_A^{m-l}\right).\nonumber
\end{align}
We now use the Schwartz inequality
\begin{align}
|\tr(\rho A^*BC)| &\leq \left[\tr(\rho A^*A)\tr(\rho C^*B^*BC)\right]^{1/2} \\
&\leq \norm{B} \left[\tr(\rho A^*A)\tr(\rho C^*C)\right]^{1/2},\nonumber
\end{align}
together with inequality \eqref{eq:inequality} to obtain
\begin{align}
\left|\frac{\text{1st term}}{\Tr(Q_A^{m}\rho (Q_A^*)^{m})^{1/2}\Tr(Q_A^{m'}\rho (Q_A^*)^{m'})^{1/2}}\right| &\leq 2\norm{A}\sum_{k=0}^{m'}\sum_{l=1}^m{m'\choose k}{m\choose l} \left(\frac{|A|}{\mu N}\right)^{k+l} \\
&\leq 2\norm{A}\exp\left(\frac{|A|m'}{\mu N}\right)\left(\exp\left(\frac{|A|m}{\mu N}\right)-1\right)\nonumber\\
&\leq 2\norm{A}\exp\left(\frac{|A|M}{\mu N}\right)\left(\exp\left(\frac{|A|M}{\mu N}\right)-1\right)\nonumber\\
&\leq 2\norm{A}\frac{16|A|}{\mu^2}\e^{\mu/16}(\e^{\mu/16}-1)\frac{M}{N},\nonumber
\end{align}
where we assumed $N\geq \frac{16|A|^2}{\mu^2}$ and $|\frac{m}{N}|\leq \frac{\mu^2}{16|A|}$. Multiplying with the ratio \eqref{eq:ratio} we obtain the desired bound.
\end{proof}
\end{lemma}

Let us now turn to the proof of the theorem. We note that the proof does not depend on the symmetry of the Liouvillian, just on the symmetry and long-range order of $\rho$, and the locality and reversibility of the dynamics.
Without loss of generality we can assume that $m,m'\geq 0$ since otherwise we merely have to exchange $O^+$ and $O^-$ and some operators with their adjoints in the proof.

By the above Lemma, we have
\begin{align}
\Delta^{(m,m')}(A,B) \simeq \frac{\omega\left( (O^-)^{m'}(O^+)^{m}(\SOp[\mcL]{A}B-A\SOp[\mcL]{B})\right)}{\omega\left((O^+)^{m} (O^-)^{m}\right)^{1/2}\omega\left((O^+)^{m'}(O^-)^{m'}\right)^{1/2}}=: \omega(\Omega^{(m,m')}(\SOp[\mcL]{A}B-A\SOp[\mcL]{B})),
\end{align}
where the $\simeq$ denotes equality up to terms that vanish in the thermodynamic limit and we have introduced the operator
\begin{equation}
\Omega^{(m,m')} := \frac{(O^-)^{m'}(O^+)^{m}}{\omega\left((O^+)^{m} (O^-)^{m}\right)^{1/2}\omega\left((O^+)^{m'}(O^-)^{m'}\right)^{1/2}}.
\end{equation}
We will now first approximate $\mcL$ by a strictly local Liouvillian $\tL$, by approximating each local term $\mcL_x$ by a term $\tL^l_x$ that is supported within the ball of radius $l$ around $x$. For each term, this introduces at most an error $c f(l)$. We collect the correcting terms in an error term $\mcR$, so that we have
\begin{equation}
\mcL = \tL + \mcR.
\end{equation}
\begin{lemma}[Approximate detailed balance] The Liouvillian $\tL$ satisfies approximate detailed balance with respect to $\omega$: For any two operators we have
\begin{align}
|\omega(\SOp[\tmcL]{A}B) - \omega(A\SOp[\tmcL]{B})| = |\omega(\SOp[\mcR]{A}B) - \omega(A \SOp[\mcR]{B})|
&\leq 2 |\Lambda | c f(l) \norm{A}\norm{B}.
\end{align}
\begin{proof}
The claim follows immediately from $|\omega(X)|\leq \norm{X}$ for any state $\omega$ and operator $X$.
\end{proof}
\end{lemma}
Remembering that $f(l)$ decays faster than any polynomial, the above Lemma shows that, even if $A$ or $B$ grow polynomially with the system size, $\tL$ is
asymptotically in detailed balance if we choose that $l$ grows at least like $L^{\alpha}$ for some $0<\alpha <1$.

Similarly, if we write $\tilde{\Delta}^{(m,m')}$ for the same quantity as $\Delta^{(m,m')}$, but where we replace $\mcL$ with $\tL$, we obtain
\begin{equation}
\left|\Delta^{(m,m')}(A,B) - \tilde{\Delta}^{(m,m')}(A,B)\right| \leq 2  \norm{A}\norm{B} |\Lambda| c f(l).
\end{equation}
In particular, if we can choose $l \propto L^{\alpha}$ for some constant $0<\alpha<1$, this error vanishes in the thermodynamic limit. We will therefore now consider $\tilde{\Delta}^{(m,m')}$ and show that it vanishes in the thermodynamic limit as long as we choose $\alpha<1/2$.

So suppose from now on that $l=L^{\alpha}$ with $\alpha<1/2$. We will again use an approximate derivation-property of the Liouvillian $\tL$ together with the fact that it is asymptotically reversible with respect to $\omega$. To do that we will denote by $\tLA$ the Liouvillian containing all terms of $\tL$ whose support has overlap with $A$. Due to the locality of $\tL$ there are at most $|\tilde{A}|\leq |A|l^d$ such terms.

The following Lemma will be, together with Lemma~\ref{lemma:koma}, the key result to prove the theorem.
This property will be particularly useful in combination with Lemma \ref{lemma:approx_derivation_app}.

\begin{lemma}[Asymptotically local derivation]\label{lemma:asym_local_derivation} Let $A$ be a local observable and let $f(l)$ grow as most like $L^{\alpha}$ with $0<\alpha<1$ with the system size. Then
\begin{equation}
\lim_{\Lambda \nearrow \Z^d} \omega(\Gamma_{\tL}(\Omega^{(m,m')},A)) = \lim_{\Lambda \nearrow \Z^d} \omega(\Gamma_{\tLA}(\Omega^{(m,m')},A)) = 0.
\end{equation}
\end{lemma}
Before we give the proof of this Lemma, we will show how it implies the main theorem. The steps are essentially the same as in the case of discrete symmetry breaking. Let $A,B$ be local operators. We first use approximate detailed balance together with the approximate derivation property to rewrite $\tilde{\Delta}^{(m,m')}(A,B)$
\begin{align}
\tilde{\Delta}^{(m,m')}(A,B)&\simeq \omega\left(\Omega^{(m,m')}\left(A\SOp[\tmcL]{B} - \SOp[\tmcL]{A}B\right)\right) \simeq \omega\left(\left(\tmcL(\Omega^{(m,m')}A)-\Omega^{(m,m')}\SOp[\tmcL]{A}\right)B\right) \\
&\simeq \omega\left(\SOp[\tmcL]{\Omega^{(m,m')}}AB\right) \simeq \omega\left(\Omega^{(m,m')}\SOp[\tmcL]{AB}\right),\nonumber
\end{align}
where again $\simeq$ denotes equality up to terms that vanish in the thermodynamic limit and where we have used approximate detailed balance in the last step. On the other hand, since $AB$ is also a local observable, we can also use the approximate derivation property to show
\begin{align}
\omega\left(\SOp[\tmcL]{\Omega^{(m,m')}}AB\right) \simeq \omega\left(\SOp[\tmcL]{\Omega^{(m,m')}AB}\right) - \omega\left(\Omega^{(m,m')}\SOp[\tmcL]{AB}\right) \simeq - \omega\left(\Omega^{(m,m')}\SOp[\tmcL]{AB}\right).
\end{align}
Combining the two estimates with $\tilde{\Delta}^{(m,m')}(A,B)\simeq \Delta^{(m,m')}(A,B)$ we therefore get
\begin{equation}
- \omega\left(\Omega^{(m,m')}\SOp[\tmcL]{AB}\right) \simeq \Delta^{(m,m')}(A,B) \simeq \omega\left(\Omega^{(m,m')}\SOp[\tmcL]{AB}\right).
\end{equation}
In other words
\begin{equation}
\lim_{\Lambda\nearrow \Z^d} \Delta^{(m,m')}(A,B) = 0.
\end{equation}

\begin{proof}(Of Lemma~\ref{lemma:asym_local_derivation}) To prove the lemma, we split up $O^+$ as $O^+=Q+R$, where $Q$ is supported on the complement of $\tilde{A}$ and $R$ collects the remaining terms. In particular this means that $\SOp[\tLA]{QX}=Q\SOp[\tLA]{X}$ for any operator $X$.
Let us also introduce the short-hand notation
\begin{equation}
Z^{(m,m')} := \omega((O^+)^{m}(O^-)^{m})^{1/2}\omega((O^+)^{m'}(O^-)^{m'})^{1/2}.
\end{equation}
We now use a binomial expansion to write
\begin{align}
\left|\omega(\Gamma_{\tLA}(\Omega^{(m,m')},A))\right| &\leq \sum_{k=0}^m\sum_{k'=0}^{m'}{m \choose k}{m' \choose k'} \left|\frac{\omega\left(\Gamma_{\tLA}((Q^\dagger)^{m'-k}Q^{m-k}(R^\dagger)^{k'}R^k,A)\right)}{Z^{(m,m')}}\right| \nonumber\\
&= \sum_{k=0}^m\sum_{k'=0}^{m'}{m \choose k}{m' \choose k'} \left|\frac{\omega\left((Q^\dagger)^{m'-k}Q^{m-k}\Gamma_{\tLA}((R^\dagger)^{k'}R^k,A)\right)}{Z^{(m,m')}}\right|.
\end{align}
But $\Gamma_{\mcL}(\one,X)=0$ for any operator $X$ and any Liouvillian $\mcL$. Therefore we can neglect the term with $k'=k=0$. Combining this
with another application of the Cauchy-Schwartz inequality we get
\begin{align}
\left|\omega(\Gamma_{\tLA}(\Omega^{(m,m')},A))\right| &\leq \sideset{}{'}\sum_{k,k'}{m \choose k}{m' \choose k'} \frac{\omega\left((Q^\dagger)^{m'-k}Q^{m'-k}\right)^{1/2}\omega\left((Q^\dagger)^{m-k}Q^{m-k}\right)^{1/2}}{Z^{(m,m')}}\norm{\Gamma_{\tLA}((R^\dagger)^{k'}R^k,A)},
\end{align}
where the primed sum omits the term $k'=k=0$. We can now use Lemma~\ref{lemma:koma} to bound the fraction as
\begin{align}
\frac{\omega\left((Q^\dagger)^{m'-k}Q^{m'-k}\right)^{1/2}\omega\left((Q^\dagger)^{m-k}Q^{m-k}\right)^{1/2}}{Z^{(m,m')}} \leq \frac{(\mu o L^d)^{-2(k+k')}}{2-\e^{\mu/8}},
\end{align}
provided that $L^d\geq \frac{16|\tilde{A}|^2}{\mu^2}$ and $|\frac{M}{L^d}|\leq \frac{\mu^2}{16|\tilde{A}|}$, where $M\geq |m|,|m'|$. Since by assumption $|\tilde{A}|\leq |A|L^{\alpha d}$, the  inequalities are fulfilled for large enough system sizes as long as $\alpha < 1/2$.
Similarly, by the locality of the Liouvillian, we can upper bound the norm-factor as
\begin{align}
\norm{\Gamma_{\tLA}((R^\dagger)^{k'}R^k,A)} \leq  3b|\tilde{A}|\norm{R}^{k+k'}\norm{A}\leq 3 b |A|\norm{A} L^{\alpha d} \left(o |A| L^{\alpha d}\right)^{k+k'}.
\end{align}
Combining the two estimates we get
\begin{align}
\left|\omega(\Gamma_{\tLA}(\Omega^{(m,m')},A))\right| &\leq \frac{3 b |A|\norm{A}}{2-\e^{\mu/8}} L^{\alpha d} \sideset{}{'}\sum_{k,k'}{m \choose k}{m' \choose k'} \left(\frac{|A| }{\mu} L^{(\alpha-1)d}\right)^{k+k'} 
\nonumber \\
&\leq \frac{3 b |A|\norm{A}}{2-\e^{\mu/8}} L^{\alpha d} \left(\exp(\frac{|A| }{\mu} M L^{(\alpha-1)d})-1\right).
\end{align}
But $L^{\alpha d}\left(\exp\left(\frac{|A| }{\mu} M L^{(\alpha-1)d}\right)-1\right)$ converges to zero as $L\rightarrow \infty$ as long as $\alpha<1/2$. This finishes the proof.
\end{proof}

\subsection{Proof of Lemma \ref{lemma:koma}}

Let $a_m:= \Tr(Q_A^{m}\rho (Q_A^*)^{m})$. We have to prove
\begin{equation}
\frac{a_m}{a_{m-1}}\geq (\mu o N)^2. 
\end{equation}
We first calculate $a_1$,
\begin{align}
a_1 &= \Tr((O^+-R_A)\rho (O^--R_A^*)) \\
&\geq \Tr(\rho O^-O^+)-2\norm{O^+R_A^*}\leq 2N o^2 |A|\nonumber\\
&=\frac{1}{2}\left[ \Tr(\rho O^+ O^-)  \Tr(\rho O^-O^+) + \Tr(\rho[O^+,O^-])\right]-2o^2 N|A|\nonumber\\
&\geq \Tr(\rho {O^{(1)}}^2) +  \Tr(\rho {O^{(2)}}^2) - 2o^2 (1+|A|)N\nonumber\\
&\geq 2o^2\mu^2 N^2\left[1 - \frac{1+|A|}{\mu^2 N} \right].\nonumber
\end{align}
Using the bound $N\geq \frac{16|A|^2}{\mu^2}$ we have
\begin{align}
1 - \frac{1+|A|}{\mu^2 N} \geq 1 - \frac{1+|A|}{16 |A|^2} \geq 1 - \frac{1}{8} > 0,
\end{align}
since $|A|\geq 1$. Therefore $a_1>0$. Next we can again use the Schwartz inequality to get
\begin{align}
(a_{m-1})^2 &\leq \Tr(\rho (Q_A^*)^{m-2}Q_A^{m-2})\Tr(\rho (Q_A^*)^{m-1}Q_AQ_A^*Q_A^{m-1}) \\
&= a_{m-2}\left\{\Tr(\rho (Q_A^*)^{m}Q_A^{m}) + \Tr(\rho (Q_A^*)^{m-1}[Q_A,Q_A^*]Q_A^{m-1})\right\}\nonumber\\
&\leq a_{m-2}\left\{a_m + 4o^2 N a_{m-1}\right\}.\nonumber
\end{align}
Assuming $a_{m-2}\neq 0,a_{m-1}\neq 0$, which is true for $m=2$, we get
\begin{align}
\frac{a_m}{a_{m-1}}\geq \frac{a_{m-1}}{a_{m-2}}-4o^2N.
\end{align}
Summing up, we obtain
\begin{align}
\frac{a_m}{a_{m-1}}&\geq a_1 - 4o^2N(m-2) \\
&\geq 2(\mu o N)^2\left[1 - \frac{1+|A|}{\mu^2 N} - \frac{2(m-2)}{\mu^2 N}\right] \nonumber\\
&\geq 2(\mu o N)^2\left[1 - \frac{1+|A|}{\mu^2 N} - \frac{2M}{\mu^2 N}\right] \nonumber\\
&\geq 2(\mu o N)^2\left[1 - \frac{1+|A|}{16|A|^2} - \frac{1}{8 |A|}\right]\nonumber \\
&\geq 2(\mu o N)^2\left[\frac{16-2-2}{16} \right] = (\mu o N)^2 \frac{3}{2} > (\mu o N)^2,\nonumber
\end{align}
where we have used $N\geq \frac{16|A|^2}{\mu^2}$, $|\frac{M}{N}|\leq \frac{\mu^2}{16|A|}$ and $|A|\geq 1$. The desired bound thus holds by induction.
Let us now lower bound the ratio
\begin{equation}
r_A^{(M)} = \left|\frac{\tr\left((O^+)^M \rho (O^-)^M\right)}{\Tr(Q_A^{M}\rho (Q_A^*)^{M})}\right|.
\end{equation}
We use a binomial expansion again to first obtain
\begin{align}
\left| \tr\left((O^+)^M \rho (O^-)^M\right) \right| &=  \left|\Tr(Q_A^{M}\rho (Q_A^*)^{M}) + \sum'_{k,l} {M\choose k}{M\choose l}\tr\left(\rho (Q_A^*)^{M-k}(R_A^*)^kQ_A^{M-l}R_A^{M-l}\right)\right| \\
&\geq |\Tr(Q_A^{M}\rho (Q_A^*)^{M})| - \sum'_{k,l} \left|{M\choose k}{M\choose l}\tr\left(\rho (Q_A^*)^{M-k}(R_A^*)^kQ_A^{M-l}R_A^{M-l}\right)\right|,\nonumber
\end{align}
where the primed sum goes over all $k,l=0,\ldots , M$ except for $k=l=0$. Using the Schwartz inequality and \eqref{eq:inequality} again we get the bound
\begin{align}
r_A^{(M)}&\geq 1-\sum'_{k,l}(|A|o)^{k+l} (\mu o N)^{-(k+l)} \\
&\geq 1 - \left[\left(1+\frac{|A|}{\mu N}\right)^{2M}-1\right]\nonumber\\
&\geq 2 - \exp(\frac{2|A|M}{\mu N}) \geq 2 - \e^{\mu/8}.\nonumber
\end{align}
Note  that, in particular, $r_A^{(M)}>0$, since $0 < \mu\leq 1$

\section{Goldstone modes}

Here we give a sketch of how to construct dissipative Goldstone modes above a symmetry-broken steady-state if the Liouvillian is symmetric and commutes with charge in the
sense of eq.~(\ref{eq:symmetry}). For simplicity we will assume that the dynamics is strictly local.
For any cubical volume $\Lambda\subseteq \Z^d$ of side-length $L$ and local region $A\subset \Lambda$, define
\begin{align}
U_{A} := \exp\left(\frac{2\pi  \mathrm{i}}{L}\sum_{x\in A} \sum_{j=1}^d x_j C_x\right).
\end{align}
The operator $U_{\Lambda}$ creates a spin-wave of wavelength $L$ on the whole volume $\Lambda$. We can then define states
\begin{equation}
\sigma_\Lambda^{(M)}(A) = \omega_\Lambda^{(M)}\left((U_\Lambda)^\dagger A U_{\Lambda}\right).
\end{equation}
For large $M$ these describe symmetry-broken states with one spin-wave excitation in each space-direction.
Now fix some local observable $A$. As $\Lambda$ increases, we can approximate $U_A$ by an operator $V_A$ that effects a spatially \emph{constant} rotation in the region $A$ in the sense that we have
\begin{align}
\norm{U_{\tilde{A}} \SOp[\mcL^\Lambda_A]{A}U_{\tilde{A}}^\dagger - V_{\tilde{A}}^\dagger \SOp[\mcL^\Lambda_A]{A}V_{\tilde{A}}} \leq 2\pi o\frac{\mathrm{diam}(A) |A|}{L} + O(1/L^2).
\end{align}
We thus obtain
\begin{align}
\left|\sigma_\Lambda^{(M)}\!\left(\SOp[\mcL^\Lambda]{A}\right) - \omega_\Lambda^{(M)}\! \left(\SOp[\mcL^\Lambda]{A}\right) \right| &\approx \left|\omega_\Lambda^{(M)} \left( \SOp[\mcL^\Lambda]{V_A^\dagger A V_A}\right)\right|   \\
&\approx 0,\nonumber
\end{align}
where $\approx$ denotes equality up to a difference of order $1/L$.
Thus time-derivatives of local observables become vanishingly small as the system-size (and wave-length) increases. The actual expectation values instead can differ arbitrarily. In particular for any $L$ the order parameter ${\bf m}(x) = (O^{(1)}_x,O^{(2)}_x)$ perfectly distinguishes the two states: Its image ${\bf m}(\Lambda)\subset \R^2$ is a single point for $\omega_\Lambda^{(M)}$ and an arbitrarily dense (as $L$ increases) circle for $\sigma_\Lambda^{(M)}$. Note that the above arguments did not rely on the reversibility assumption, but only on the locality and symmetry of the Liouvillian under the action of a locally generated symmetry.

\end{document}